\ifdefined\spellcheck
\documentclass{spellchecking}
\else
\documentclass[conference]{IEEEtran}
\fi
\IEEEoverridecommandlockouts
\usepackage{cite}
\usepackage{amsmath,amssymb,amsfonts}
\usepackage{algorithmic}
\usepackage{graphicx}
\usepackage{textcomp}
\usepackage{xcolor}

\usepackage[absolute]{textpos}
\usepackage{scalerel}
\usepackage{tikz}
\usetikzlibrary{svg.path}

\definecolor{orcidlogocol}{HTML}{A6CE39}
\tikzset{
  orcidlogo/.pic={
    \fill[orcidlogocol] svg{M256,128c0,70.7-57.3,128-128,128C57.3,256,0,198.7,0,128C0,57.3,57.3,0,128,0C198.7,0,256,57.3,256,128z};
    \fill[white] svg{M86.3,186.2H70.9V79.1h15.4v48.4V186.2z}
                 svg{M108.9,79.1h41.6c39.6,0,57,28.3,57,53.6c0,27.5-21.5,53.6-56.8,53.6h-41.8V79.1z M124.3,172.4h24.5c34.9,0,42.9-26.5,42.9-39.7c0-21.5-13.7-39.7-43.7-39.7h-23.7V172.4z}
                 svg{M88.7,56.8c0,5.5-4.5,10.1-10.1,10.1c-5.6,0-10.1-4.6-10.1-10.1c0-5.6,4.5-10.1,10.1-10.1C84.2,46.7,88.7,51.3,88.7,56.8z};
  }
}

\newcommand\orcidicon[1]{\href{https://orcid.org/#1}{\mbox{\scalerel*{
\begin{tikzpicture}[yscale=-1,transform shape]
\pic{orcidlogo};
\end{tikzpicture}
}{|}}}}

\ifdefined\spellcheck
    \usepackage{subfig}
\else
    \ifCLASSOPTIONcompsoc
        \usepackage[caption=false,font=normalsize,labelfont=sf,textfont=sf]{subfig}
    \else
        \usepackage[caption=false,font=footnotesize]{subfig}
    \fi
\fi

\usepackage[]{url}
\usepackage{xspace}
\usepackage[binary-units=true]{siunitx}
\usepackage[nolist]{acronym}
\usepackage{dblfloatfix}
\usepackage{accents}
\usepackage[hidelinks]{hyperref}
\usepackage{cleveref}

\allowdisplaybreaks

\usepackage{microtype}
\usepackage[moderate, mathspacing=normal]{savetrees}

\begin{acronym}[Bash]
	\acro{CDC}{colorless, directionless, contentionless}
	\acro{CPT}{Coherent Pluggable Transceiver}
	\acro{CET}{Coherent Elastic Transponder}
	\acro{EDFA}{Erbium-Doped Fiber Amplifier}
	\acro{FEC}{Forward Error Correction}
	\acro{QoS}{Quality of Service}
	\acro{SDN}{Software-Defined Networking}
	\acro{HbH}{Hop-by-Hop}
	\acro{ILP}{Integer Linear Program}
	\acro{IQR}{Interquartile Range}
	\acro{PCS}{Probabilistic Constellation Shaping}
	\acro{SLA}{Service Level Agreement}
	\acro{DSP}{Digital Signal Processing}
	\acro{DCI}{Data Center Interconnect}
	\acro{MSA}{Multi-Source Agreement}
	\acro{OXC}{Optical Cross-Connect}
	\acro{O-E-O}{Optical-Electrical-Optical}
	\acro{ROADM}{Reconfigurable Optical Add-Drop Multiplexer}
	\acro{WDM}{Wavelength-Division Multiplexing}
	\acro{IA}{Implementation Agreement}
	\acro{IBN}{Intent-Based Networking}
	\acro{NBI}{Northbound Interface}
	\acro{EWBI}{East\slash{}Westbound Interface}
	\acro{ML}{Machine Learning}
	\acro{E2E}{end-to-end}
	\acro{RAN}{Radio Access Network}
	\acro{SFC}{Service Function Chain}
	\acro{DC}{Data Center}
	\acro{IPoDWDM}{IP over DWDM}
	\acro{IaaS}{Infrastructure-as-a-Service}
	\acro{OIF}{Optical Internetworking Forum}
	\acro{DCO}{Digital Coherent Optics}
	\acro{BGP}{Border Gateway Protocol}
	\acro{DAG}{Directed Acyclic Graph}
	\acro{RMSA}{Routing, Modulation, and Spectrum Assignment}
	\acro{RSA}{Routing and Spectrum Assignment}
	\acro{MD}{multi-domain}
	\acro{AS}{Autonomous System}
	\acro{MCMC}{Markov Chain Monte Carlo}
	\acro{PPL}{Probabilistic Programming Language}
    \acro{PDF}{probability density function}
    \acro{NUTS}{No-U-Turn Sampler}
    \acro{MTTF}{Mean Time To Failure}
    \acro{MTTR}{Mean Time To Repair}
    \acro{LHS}{Left-Hand Side}
    \acro{RHS}{Right-Hand Side}
    \acro{RV}{Random Variable}
    \acro{ESS}{effective sample size}
    \acro{PSRF}{potential scale reduction factor}
    \acro{PCI}{percentile central interval}
    \acro{RMSE}{Root Mean Square Error}
    \acro{HDI}{Highest Density Interval}
    \acro{DAG}{Directed Acyclic Graph}
    \acro{SAP}{Shortest Available Path}
    \acro{JML}{Joint Multilayer}
    \acro{LDJML}{Latency-Driven Joint Multilayer}
    \acro{NFV}{Network Function Virtualization}
    \acro{VNF}{Virtual Network Function}
    \acro{SFC}{Service Function Chain}
\end{acronym}

\newcommand*{\figref}[1]{Fig.~\ref{#1}}
\newcommand*{\secref}[1]{Section~\ref{#1}}

\newlength{\dhatheight}

\crefrangelabelformat{equation}{(#3#1#4--#5#2#6)}
\crefname{equation}{}{}

\def\BibTeX{{\rm B\kern-.05em{\sc i\kern-.025em b}\kern-.08em
    T\kern-.1667em\lower.7ex\hbox{E}\kern-.125emX}}

\begin{document}
\title{Grooming Connectivity Intents in IP-Optical Networks Using Directed Acyclic Graphs
\thanks{This work has been performed in the framework of the CELTIC-NEXT EUREKA project AI-NET-ANTILLAS (Project ID C2019/3-3), and it is partly funded by the German BMBF (Project ID 16KIS1312).}
}

\author{
\IEEEauthorblockN{Filippos Christou  \orcidicon{0000-0001-5181-378X}}
\IEEEauthorblockA{\textit{Institute of Communication Networks}\\
\textit{and Computer Engineering (IKR)} \\
\textit{University of Stuttgart}\\
Stuttgart, Germany \\
filippos.christou@ikr.uni-stuttgart.de} 
\and
\IEEEauthorblockN{Andreas Kirst\"adter}
\IEEEauthorblockA{\textit{Institute of Communication Networks}\\
\textit{and Computer Engineering (IKR)} \\
\textit{University of Stuttgart}\\
Stuttgart, Germany \\
andreas.kirstaedter@ikr.uni-stuttgart.de}
}

\maketitle

\begin{textblock}{13.5}(1,0.15)
    {\color{black}
    © 2023 IEEE. Personal use of this material is permitted. Permission from IEEE must be
    obtained for all other uses, in any current or future media, including
    reprinting/republishing this material for advertising or promotional purposes, creating new
    collective works, for resale or redistribution to servers or lists, or reuse of any copyrighted
    component of this work in other works.
    }
\end{textblock}

\begin{abstract}
    During the last few years, there have been concentrated efforts toward intent-driven networking.
    While relying upon \ac{SDN}, \ac{IBN} pushes the frontiers of efficient networking by decoupling the intentions of a network operator (i.e., what is desired to be done) from the implementation (i.e., how is it achieved).
    The advantages of such a paradigm have long been argued and include, but are not limited to, the reduction of human errors, reduced  expertise requirements among operator personnel, and faster business plan adaptation.
    In previous work, we have shown how incorporating \ac{IBN} in multi-domain networks can have a significantly positive impact as it can enable decentralized operation, accountability, and confidentiality.
    The pillar of our previous contribution is the compilation of intents using system-generated intent trees.
    In this work, we extend the architecture to enable grooming among the user intents.
    Therefore, separate intents can now end up using the same network resources.
    While this makes the intent system reasonably more complex, it indisputably improves resource allocation.
    To represent the intent relationships of the newly enhanced architecture, we use \acp{DAG}.
%    Furthermore, we employ an advanced technique \cite{GKAMAS} to jointly solve the \ac{RMSA} problem, which we appropriately adapt to be used for the intent compilation and highlight the differences from the predecessor.
    Furthermore, we appropriately adapt an advanced established technique from the literature to solve the \ac{RMSA} problem for the intent compilation.
    We demonstrate a realistic scenario in which we evaluate our architecture and the intent compilation strategy.
    Our current approach successfully consolidates the advantages of having an intent-driven architecture and, at the same time, flexibly choosing among advanced resource allocation techniques.
%    This work constitutes a realistic implementation of an IBN framework for IP-Optical networks, which was evaluated using simulation methods.
\end{abstract}


\acresetall

\section{Introduction}

With the dramatic growth of data traffic in IP-optical networks, efficient and scalable network control and management solutions have become increasingly critical.
Intent-driven networking has emerged as a promising approach to simplify these tasks by allowing operators to express high-level network objectives as intents.
Connectivity intents define the desired end-to-end connectivity between network nodes and are used to automatically configure the underlying network infrastructure. 
The \ac{IBN} framework responsible for this automatic implementation is logically placed on top of the \ac{SDN} controller and carries the network logic.
The \ac{SDN} controller then gets exclusively dedicated to facilitating the communication between the \ac{IBN} framework and the network devices; it receives a role similar to a device driver in computer systems.
Thus, the operator will send connectivity intents to the \ac{IBN} framework.
The \ac{IBN} framework compiles the intents to an implementation, which is then forwarded to the \ac{SDN} controller to be installed into the devices.

One of the critical design choices in operating an \ac{IBN} framework is choosing the compilation algorithm.
The compilation algorithm receives an abstract high-level network objective and automatically outputs an implementation that can be realized in the network at a specific time.
Depending on the objective's nature, different algorithms must be invoked.
For example, \ac{NFV} intents could be addressed using algorithms for \acp{VNF} placement and \ac{SFC} deployment \cite{2018BoYi, 2019Miladic}.
Since we consider IP-optical connectivity intents in this work, the serving algorithmic family is that of \ac{RMSA} \cite{RMSA}.

Ideally, we would like to reuse any of the algorithms from the literature in the intent-driven environment.
This can be done effortlessly using a one-step compilation procedure and independently invoking the preferred algorithm when new intents must be compiled.
The resulting interoperability with legacy algorithms would allow some of the advantages of \ac{IBN}, like the holistic view and the built-in architecture for the intent lifecycle.
However, a mere "re-branding" of the algorithms without truly adapting them to the intent-driven architecture would require re-implementation of procedures like intent monitoring (making sure that installed intents remain successfully installed), intent conflicts resolution (when more than one intent requires the same resources), and intent re-provisioning (in case of network failures) each time according to the legacy algorithm chosen.

In our previous work \cite{CHRISTOUCNSM}, we have outlined a series of advantages that can be gained by having a multi-step intent compilation approach.
Specifically, we have concluded that hierarchical system-generated intent structures, like intent trees, can offer efficient decentralized coordination of multi-domain IP-optical networks.
Although possibly requiring more effort to bind to legacy algorithms, this approach promises flexible and scalable communication mechanisms, confidentiality, and accountability.
However, there is no inherent possibility to use algorithms that leverage grooming as long as intent trees are used for the representation of the multi-step intent compilation.
IP-optical traffic grooming \cite{2019Miladic} is a popular technique that packs several demands into the same optical spectrum channel, thus increasing resource utilization.

This paper fills this gap by developing a new approach that permits IP-optical grooming by substituting the intent trees with an intent \ac{DAG}.
As a result, we reinvent an intent-driven framework that inherits all the benefits from its predecessor and can successfully integrate any grooming-enabled \ac{RMSA} algorithm.
To demonstrate this universality, we adopt an established \ac{RMSA} algorithm from the literature \cite{GKAMAS}.
We also proceed to slight modifications to prioritize low-latency paths, showing the algorithm's preserved flexibility in the intent-driven environment.
We validate the operation of the architecture by simulating a realistic scenario with well-known results.
We limit our focus on single-domain scenarios to better focus on the task at hand.

In the following section, we introduce the new architecture using an intent \ac{DAG} as opposed to the one using intent trees.
\secref{secalg} describes the compilation algorithms and the integration steps needed for \cite{GKAMAS}.
In \secref{seceval}, we evaluate our architecture in a realistic scenario seeking to reproduce the expected well-known results.
In \secref{secconc}, we conclude the paper and disclose future directions.

\section{Architecture}\label{secarch}

\begin{figure}[b]
    \vspace{-15pt}
	\centerline{\includegraphics[width=1.0\linewidth]{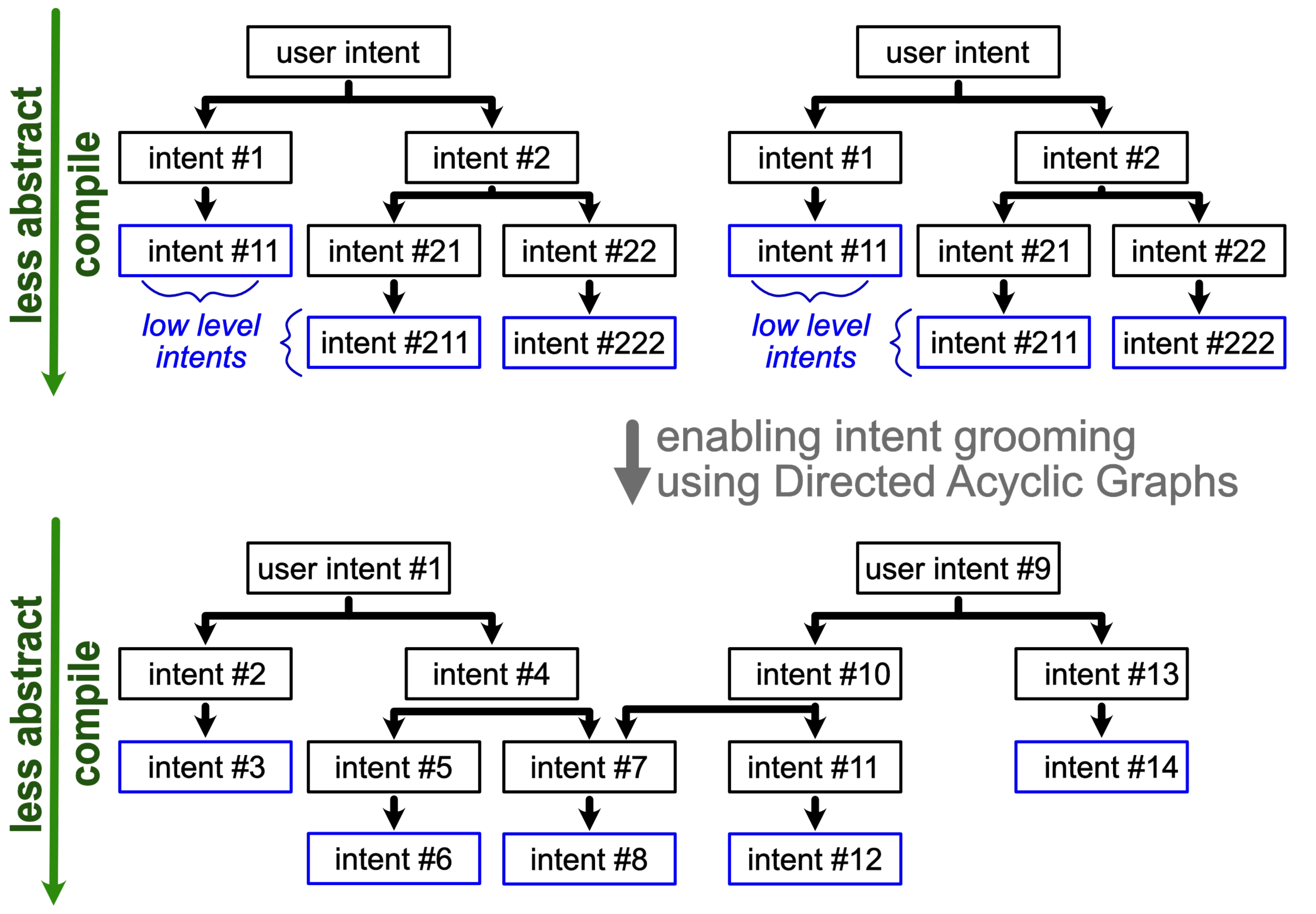}}
    \caption{Comparing intent trees and intent \ac{DAG} \label{fig:strat}}
\end{figure}

\begin{figure}[b]
    \vspace{-14pt}
	\centerline{\includegraphics[width=1.0\linewidth]{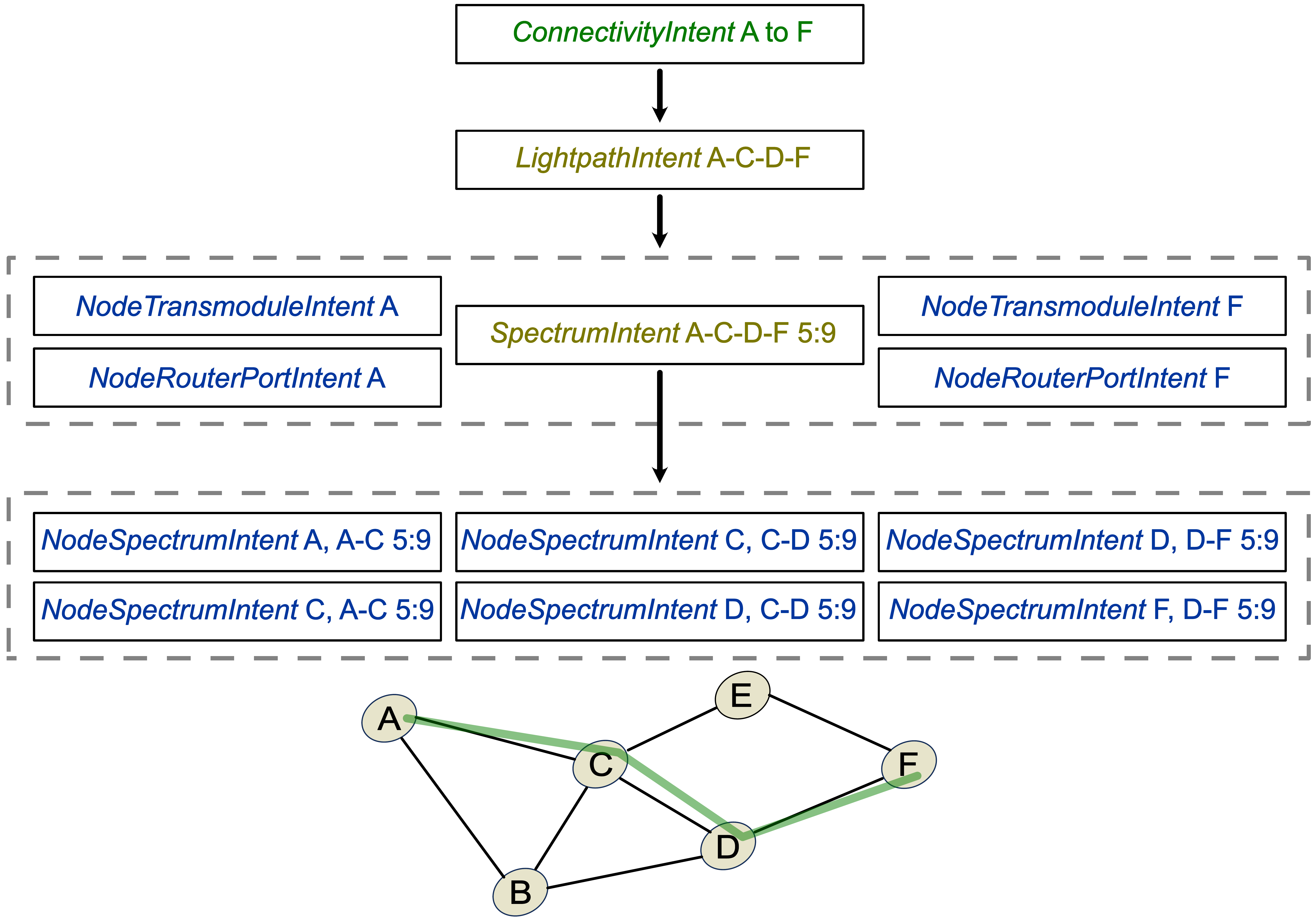}}
    \caption{Example of a compiled intent.
    The connection from node A to node F is implemented with one lightpath covering the nodes A-C-D-F and using the spectrum slots 5,6,7,8,9 along the involved fibers. \label{fig:intentdagexample}}
\end{figure}

\begin{figure*}[b]
    \vspace{-15pt}
	\centerline{\includegraphics[width=1\linewidth]{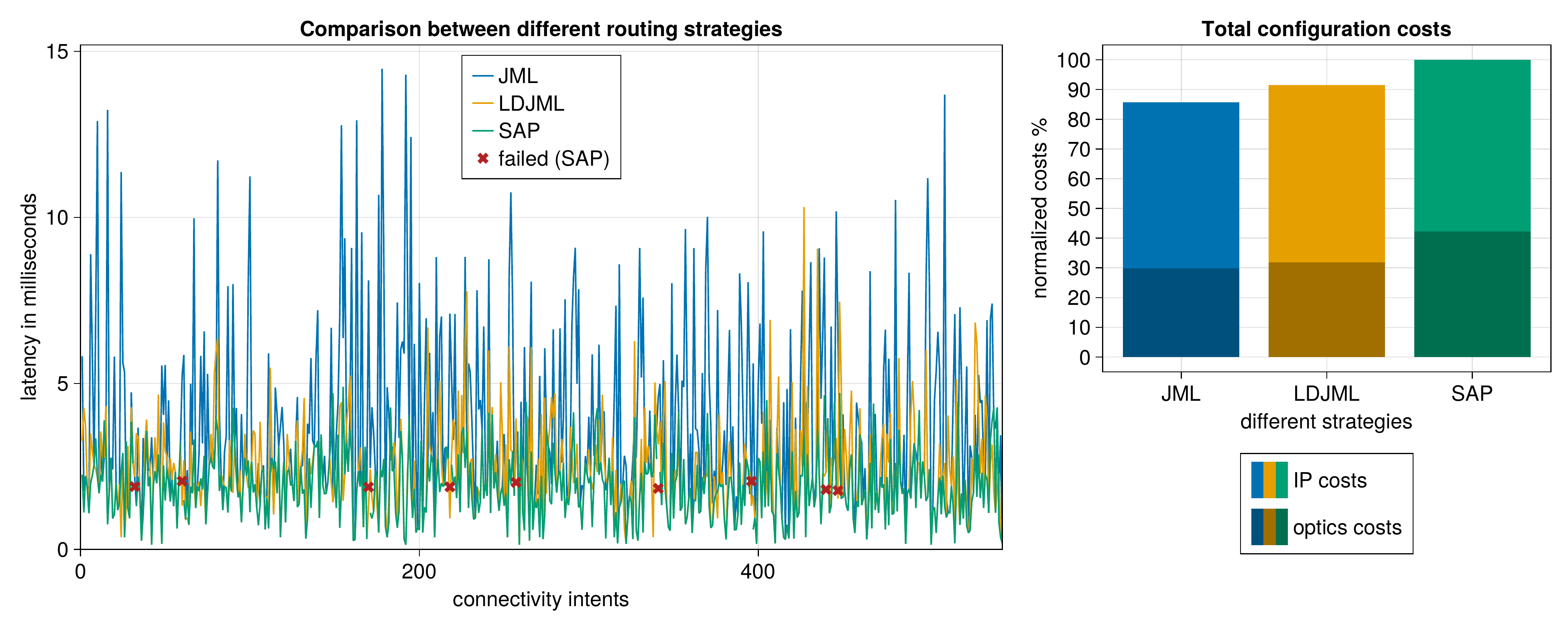}}
    \caption{Single simulation comparison. On the right, the IP and optics costs are the same as the attributes $P_p$ and $C_p$ in the path cost vector. \label{fig:evalsingleseed}}
    \vspace{-5pt}
\end{figure*}
In this section, we will first briefly revisit the architecture of \cite{CHRISTOUCNSM} using intent trees and then contrast it with the novel version using an intent \ac{DAG}.

\subsection{Intent Trees}
Intent compilation using intent trees recursively generates child intents from higher-level intents.
The intent tree is a hierarchical representation of the network intent, where the depth corresponds to a different level of abstraction. 
The tree's root represents the high-level user intent, while the leaf nodes represent the low-level intents which are the needed device configurations.

The top of \figref{fig:strat} illustrates two different intent trees corresponding to two different user intents.
The generation of the intent tree is specific to the compilation algorithm chosen.
One-step compilation algorithms are still possible; in that case, the tree would be two levels deep and be composed exclusively of the root user intent and its low-level intents children.

It is important to notice that separate user intents define separate intent trees, and there is no way of merging them.
As a result, the resources reserved by the low-level intents in one intent tree cannot be jointly used by another, rendering support for grooming impossible.

\subsection{Intent \ac{DAG}}

To reuse resources (i.e., low-level intents) across different user intents, we must allow some intent nodes to have several parents, leading us to use a \ac{DAG}.
Furthermore, \acp{DAG} preserve the direction the same way it exists in a tree and avoid unwanted cyclic dependencies.
In contrast with the intent trees, where the number of intent trees directly depends on the number of user intents, the intent \ac{DAG} is unique per \ac{IBN} framework and includes all user intents.
However, the intent \ac{DAG} might be disconnected (when no grooming is done), and then practically several partial \acp{DAG} will appear.
The bottom of \figref{fig:strat} illustrates how an intent \ac{DAG} could modify the intent structure.
A unified indexing is also necessary to combine all the intent trees.

The following network intents were defined to solve the grooming-enabled \ac{RMSA} problem.
\begin{itemize}
    \item \emph{LightpathIntent} is an intent that defines a lightpath across some network nodes.
        These intents are the grooming points for IP-optical networks. 
        If our IP-optical intent system were applied to the bottom of \figref{fig:strat}, then \linebreak intent \#7 would be a \emph{LightpathIntent}.
    \item \emph{SpectrumIntent} is an intent that defines the spectrum requirements of a parent \emph{LightpathIntent}.
    \item \emph{NodeTransmoduleIntent} is a low-level intent that requires using a particular transmission module in a multilayer node.
    \item \emph{NodeRouterPortIntent} is a low-level intent that requires using a port in the router of a multilayer node.
    \item \emph{NodeSpectrumIntent} is a low-level intent that requires the reservation of some spectrum slots in a link connected to a multilayer node.
\end{itemize}

\figref{fig:intentdagexample} shows an example of a compiled intent involving all the aforementioned intents.

\section{Intent Compilation Algorithms}\label{secalg}
On top of the intent \ac{DAG} architecture, any grooming-enabled \ac{RMSA} algorithm could be implemented.
In this section, we will look closer at three of such algorithms.

\subsection{Shortest Available Path}
The \ac{SAP} is the simplest algorithm, also used in \cite{CHRISTOUCNSM}, and is treated as the baseline.
It is an \ac{RMSA} algorithm that first solves routing using k-shortest-path and then the spectrum assignment using first-fit \cite{firstfit}.
If the path is unavailable, the next shortest path is considered.
Using algorithms without grooming like this makes no difference whether we have intent trees or an intent \ac{DAG}.

\subsection{Joint Multilayer}
The \ac{JML} algorithm has been proposed in \cite{GKAMAS}.
It is a complex algorithm that can be used in an online fashion and is composed of the following steps:
\begin{itemize}
    \item Create a directed multilayer multigraph.
\end{itemize}
The layers of the graph correspond to the optical and electrical views.
Each topology node is converted to a two-layered node composed of an IP router vertex and an \ac{OXC} vertex.
The edges in the electrical layer signify the established lightpaths (virtual links), and the edges in the physical layer are the fibers.
Inter-layer links are the transmission modules connecting the \ac{OXC} with the IP router and vice versa.
Since more than one transmission module might be available, a multigraph is needed to accommodate several edges between the same vertices.

\begin{itemize}
    \item Calculate the cost vector for all edges.
\end{itemize}
Each edge $e$ in the multilayer multigraph is described by a vector $(D_e, C_e, P_e, \bar{H}_e, F_e, \bar{W}_e, T_e, I_e, L_e)$.
$D_e$ is the  distance covered by the last transmission module, i.e., since the last regeneration.
$C_e$ is the cost of the transmission module.
$P_e$ is the cost of the router port. 
$\bar{H}_e$ is the vector of transmission module mode tuples $[(r_1, d_1, b_1), (r_2, d_2, b_2), ...]$, where $r_i, d_i, b_i$ is the transmission rate, the optical reach, and the spectrum slot requirements, respectively.
$F_e$ is a boolean variable specifying whether the link is virtual.
$\bar{W}_e$ is the boolean vector indicating the availability of the spectrum slots.
The last three components, $T_e$, $I_e$, and $L_e$, are new additions needed for the adaptation as an intent \ac{DAG} compilation algorithm.
${T_e}$ is the type of link (virtual, optical, optical-to-virtual, or virtual-to-optical) and is used because different actions are needed based on the link type.
$I_e$ is the index of the intent \ac{DAG} node if the (virtual) link corresponds to an already established \emph{LightpathIntent}.
$L_e$ is the physical link length.
Several link cost vectors can be added to create a path cost vector with similar components $(D_p, C_p, P_p, \bar{H}_p, \bar{W}_p, I_p, L_p, \bar{R}_p, \bar{p})$, where $\bar{R}_p$ are the transmission modules chosen along the way and $\bar{p}$ is the path in the multilayer multigraph.

\begin{itemize}
    \item Obtain non-dominated paths.
\end{itemize}
Here, an algorithm generates a series of candidate multilayer paths.
A multilayer path comprises a series of optical, virtual, optical-to-virtual, and virtual-to-optical links in the correct logical order.
This collection excludes the paths whose path cost vector is categorically worse than others, i.e., they are dominated by other paths.
Such comparison can only happen between paths of the same source and destination.
A path $p_1$ is categorically better (i.e., dominates) than a path $p_2$ if

\vspace{-9pt}
\begin{align}
    &D_{p_1} \leq D_{p_2} \text{ and } C_{p_1} + P_{p_1} \leq C_{p_2} + P_{p_2} \text{ and } F_{p_1} \leq F_{p_2} \nonumber \\
    &\text{ and } \max_{\text{rate}}\bar{R}_{p_1} \geq \max_{\text{rate}}\bar{R}_{p_2} \text{ and } \bar{W}_{p_1} \geq \bar{W}_{p_2} \nonumber
\end{align}

During the non-dominated paths generation, care is being taken to only output valid multilayer paths, which respect the limitations imposed by the port rate, optical reach, and bandwidth requirements in $\bar{H_p}$.
Paths picked in this step serve as candidate paths for the next step.

\begin{itemize}
    \item Select the \emph{winner path}.
\end{itemize}
An optimization function is chosen based on the preferences, and a path is selected from all the candidates that minimize this function.
The legacy optimization function the \ac{JML} algorithm uses aggregates all electrical and optical costs $C_p + P_p$.
As a result, the path is selected that minimizes these costs at best.
\begin{itemize}
    \item Allocate spectrum slots.
\end{itemize}
After choosing the winner path, the spectrum allocation is conducted with the preferred algorithm.
The legacy algorithm uses the first fit.

Following are the extra steps introduced in the procedure for the adaptation to the intent \ac{DAG} compilation.
\begin{itemize}
    \item Break solution into predefined intents.
\end{itemize}
Once the algorithm yields an output path with all the needed resource allocations, this information is mapped into the intents defined in \secref{secarch}.
The result will be similar to \figref{fig:intentdagexample}.
\begin{itemize}
    \item Attach intents to the intent \ac{DAG}.
\end{itemize}
%This is the final step of the compilation algorithm, where the newly generated nodes are added to the intent \ac{DAG} as descendants of the user intent.
In the final step of the compilation algorithm, the newly generated nodes are added to the intent \ac{DAG} as descendants of the user intent.
In case of no grooming, all intents involved will create new child intents.
If grooming takes place, not all involved intents will be added to the \ac{DAG}, but instead, an edge will be created between the involved intents and a \emph{LightpathIntent} already existing in the \ac{DAG} identified by the $I_e$ attribute.

\subsection{Latency-Driven Joint Multilayer}
After being ported into the intent-based regime, the adapted algorithm \ac{JML} should remain easy to modify as designed by the original authors of \cite{GKAMAS}.
Indeed, we can easily modify the optimization function to adjust the algorithm to focus on finding the shortest paths, rendering the intent-driven approach equally flexible.
For this reason, we had to introduce the extra attribute $L_p$ to quantify the physical length of the path, which now needs to be minimized.
With a lower priority and in case of a tie between paths, we fall back to the previous objective function $C_p + P_p$.
We called this slight variation \ac{LDJML}, which is further used during the evaluation to attain more variety of use cases.

\begin{figure}[t]
	\centerline{\includegraphics[width=1\linewidth]{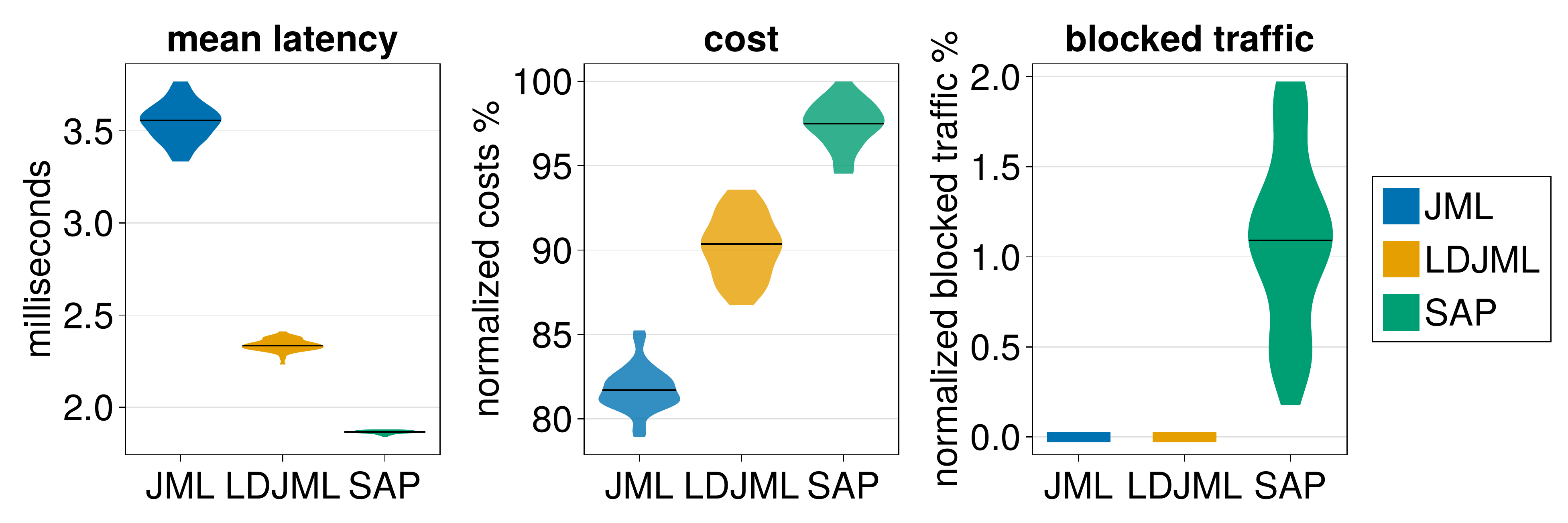}}
%    \vspace{-10pt}
    \caption{Multiple simulations comparison. The black line in the violin plots is the median. \label{fig:evalmultiseed}}
    \vspace{-15pt}
\end{figure}

\section{Evaluation}\label{seceval}
In this section, we will evaluate the presented architecture.
The results should not invoke surprise, as the core algorithms used are well-known and should yield the expected results.
This section has the role of proof of concept, where we validate that the adapted compilation algorithm \ac{JML} and its slight variation \ac{LDJML} operate as expected.

For the simulation, we used the Nobel-Germany topology from \cite{SNDlib_ref}.
The demand matrix is generated using a truncated normal distribution for every node pair, aggregating to circa \SI{62}{Tbps} for the whole network.
Each entry on the demand matrix is then used to issue a connectivity intent.
The cost and network equipment model is derived from \cite{2022ChristouTowards}.

\subsection{Single Simulation}
\figref{fig:evalsingleseed} shows the results of running a single simulation for each compilation algorithm presented in \secref{secalg}.
On the left, we can see the latency for every connectivity intent for all compilation algorithms.
\ac{SAP} generally delivers the lowest latency, \ac{JML} the highest, and \ac{LDJML} is, as designed, in the middle.
However, \ac{SAP} does not leverage grooming, and \SI{62}{Tbps} are already too much to accommodate with such a naive approach.
As a result, this leads to certain connectivity intents being blocked. 
Blocking does not happen for \ac{JML} and \ac{LDJML}, which both leverage grooming and thus manage their resources more efficiently.

We witness similar results also regarding the cost of the three compilation algorithms.
\ac{SAP} presents the highest costs as it allocates more resources for every new connectivity intent.
\ac{JML} presents the lowest costs as it is the objective of the optimization function used.
\ac{LDJML} again stands in the middle as it uses grooming, which helps reduce the costs, but also tries to minimize latency with a higher priority.

\subsection{Multiple Simulations}
To achieve more confidence in the results, we repeat the simulation 40 times using different seeds for the random generation of the demand matrix.
\figref{fig:evalmultiseed} confirms the same results as described previously.
\ac{LDJML} always stands in the middle between \ac{JML} and \ac{SAP}, with \ac{SAP} having the lowest latency, highest cost, and some blocking and \ac{JML} having the highest latency, lowest cost, and no blocking.
\ac{LDJML} also shows no blocking due to the grooming capabilities enabled by the intent \ac{DAG} compilation design.

\section{Conclusion} \label{secconc}
In this paper, we designed a modular architecture for intent compilation algorithms.
We used an intent \ac{DAG} to represent the realization of any grooming-enabled \ac{RMSA} algorithm.
Using the proposed architecture, any related algorithm can be adapted while also preserving the possibility for further modifications.
This work enables the migration of legacy algorithms to the intent-based regime for single-domain scenarios.
We demonstrated the validity of our approach by making simulations using three different intent compilation algorithms and confirming the well-known results.
As new, exciting benefits are still to be gained for decentralized multi-domain operation, future work will focus on expanding and analyzing the presence of intent \acp{DAG} in such use cases.

\bibliography{IEEEabrv,references.bib}

\end{document}